\documentclass{article}
\usepackage{authblk}
\usepackage{graphicx}
\usepackage{dcolumn}
\usepackage{cite}
\usepackage{amsmath}
\usepackage[numbers]{natbib}
\begin{document}

\title{Exotic Coherent Structures and Their Collisional Dynamics in a  (3+1) dimensional Bogoyavlensky-Konopelchenko Equation}

\author[1]{C. Senthil Kumar}
\author[2*]{R.Radha}

\affil[1]{Department of Physics, Vinayaka Mission’s Kirupananda Variyar Engineering College, Vinayaka Mission’s Research Foundation (DU), Salem  636 308, India}
\affil[2]{Centre for Nonlinear Science (CeNSc), Post-Graduate and Research Department of Physics, Government College for Women (Autonomous), Kumbakonam 612 001, India,}
\affil[*]{Correspondong Author: vittal.cnls@gmail.com}

\maketitle


\begin{abstract}
In this paper, we analyse the (3+1) dimensional Bogoyavlensky - Konopelchenko equation. Using Painlev\'e Truncation approach, we have constructed solutions in terms of lower dimensional arbitrary functions of space and time.  By suitably harnessing the arbitrary functions present in the solution, we have generated physically interesting solutions like periodic solutions, kinks, linear rogue waves, line lumps, dipole lumps  and hybrid dromions.  It is interesting to note that unlike in (2+1) dimensional nonlinear partial differential equations, the line lumps interact and undergo elastic collision without  exchange of energy which is confirmed by the asymptotic analysis.   The hybrid dromions are also found to retain their amplitudes during interaction undergoing elastic collision. The highlight of the results  is that one also observes the two nonparallel ghost solitons  as well whose intersection gives rise to hybrid dromions, a phenomenon not witnessed in (2+1) dimensions.
\end{abstract}


\maketitle
\newpage 
\section{Introduction}
 The identification of nonlinear excitations  like solitons, rogue waves, lumps, dromions  etc, which find applications in various branches of science  such as nonlinear optics\cite{Boyd08}, Bose-Einstein condensates\cite{Radha15}, fluid dynamics\cite{Apel07}, plasma physics\cite{Kono10} keeps  the investigation of nonlinear dynamical systems alive even today. 
  Rogue waves which are also known as Freak waves occur in deep ocean \cite{Kharif03,Zakharov06,Dysthe08,Osborne09}.  They arise from nowhere and disappear with no trace. Rogue waves are identified in various fields \cite{Onorato13} like hydrodynamics \cite{Chabchoub11}, nonlinear optics \cite{Solli07,Solli08,Kibler10}, Bose-
Einstein condensates \cite{Yu09,Yan10} and plasma physics \cite{Bailung11}. 
Another interesting class of solutions are lumps \cite{Ablowitz78,Estevez07}. These solutions decay algebraically and do not interact with each other. The behaviour of the above solutions have been extensively analysed in  (2+1) dimensions. A recent study shows the dynamics of lump molecules in fluid systems \cite{Zhao22}. There exists another fascinating class of solutions called "dromions" which are exponentially localized \cite{Boiti88,Fokas90,Radha94,Radha95,Lou95}. They arise by virtue of the intersection of two nonparallel ghost solitons.    Dromions have been  observed in electron-acoustic wave in space observation in the polar cap regions \cite{Ghosh00} and dust acoustic dromions have been studied in magnetized plasma \cite{Saini16}. It should be pointed out that the above exotic structures have been identified only in (2+1) dimensional nonlinear partial differential equations(pdes) and  their counterparts in (3+1) dimensions continue to be elusive. 

 Several methodologies  like inverse scattering transform\cite{Gardner67}, Hirota bilinearization method \cite{Hirota73,Raut23},  Darboux Transformation\cite{Yuan24}, B\"acklund transformation \cite{lulu22,Gao24-2,Zhou24}, Painlev\'e Analysis \cite{Gao23}, Lax pair \cite{Gao24-5}, Similarity reduction method \cite{Gao25,Gao24-1,Gao24-3,Gao24-4,Xiao24} have been used to study the above nonlinear partial differential equations.  A new analytical method, namely Painlev\'e truncation approach has been  formulated by the authors \cite{Senthil05,Senthil09:1,Senthil09:2} to study the (2+1) dimensional nonlinear partial differential equations and  construct physically interesting solutions. In this paper, we focus our attention on the extraction of such solutions in (3+1) dimensional nonlinear pdes.

In this paper, we consider the physically interesting (3+1) dimensional Bogoyavlenky - Konopelchenko equation \cite{Abdul23}. This equation describes the three dimensional interactions of Riemann waves in nonlinear media. Abdul et al have established the Painlev\'e integrability and constructed solutions using Hirota bilinear method \cite{Abdul23}.  However, it should be pointed out that the nature of the solutions and their dynamics have not yet been analyzed. In particular, the  counterparts of dromions and lumps in (3+1) dimensions  and their behaviour have not yet been understood which is going to be the focal point of the  paper. In this connection, we employ Truncated Painlev\'e approach \cite{Senthil05,Senthil09:1,Senthil09:2} to construct its solutions in terms of lower dimensional arbitrary functions of space and time.  

The plan of the paper is as follows. In section 2, we analyse the (3+1) dimensional Bogoyovlensky-Konopelchenko equation using Painlev\'e Truncation Approach and construct its solution in terms of lower dimensional arbitrary functions. In section 3, we construct physically interesting solutions like periodic waves, kinks, linear rogue waves, line and dipole lumps and hybrid dromions.  In addition, we have generalized the lump solution and constructed  N lump solution.  The elastic collision of the lumps is validated by the asymptotic analysis. We have also studied the collisional dynamics of dromions. In section 4, we conclude the results.

\section{(3+1)dimensional Bogoyavlensky-Konopelchenko Equation}
(3+1) dimensional Bogoyavlensky-Konopelchenko Equation is given by
\begin{subequations}
\begin{eqnarray}
u_{xt}+u_{yt}+u_{xxxx}+u_{xxxy}+6u_xu_{xx}+3u_x u_{xy}+3 u_{xx}u_y + \alpha u_{xy} \nonumber \\
+ \beta u_{xz} + \beta u_{yz} + \gamma_1 u_{xx} + \gamma_2 u_{yy} = 0,\label{BK}
\end{eqnarray}  
\end{subequations}
where $u$ is the field variable and  is a function of $x,y,z,t$. $\alpha, \beta, \gamma_1, \gamma_2$ are constant parameters.
\vspace{0.25cm}
\newline{\bf Theorem:} For the above (3+1) dimensional Bogoyavlensky-Konopelchenko Equation, solution can be written in a closed form as,
\begin{equation}
u = \frac{2}{x+\xi(y)} + u_1(z,t), \label{sol}
\end{equation}
where $\xi(y)$ and $u_1(z,t)$ are arbitrary functions in the indicated variables.
\vspace{0.25cm}
\newline {\bf Proof:}
We express the variable $u$ in terms of the Laurent series, 
\begin{equation}
u=\sum_{j=0}^{\infty} u_j \phi^{j-1},
\end{equation}
where $\phi$ is the noncharacteristic singular manifold and is a function of $x,y,z,t$. We have also confirmed the Painlev\'e integrability of Eq.(\ref{BK}) using reduced manifold $\phi=x+\psi(y,z,t)$ as mentioned in \cite{Abdul23}. To employ Painlev\'e truncation approach, we truncate the above Laurent series at the constant level term to give
\begin{equation}
u=\frac{u_0}{\phi}+u_1. \label{transf}
\end{equation}
Now, assuming the seed solution as
\begin{equation}
u_1=u_1(z,t).\label{seed}
\end{equation}
We now substitute the transformations  given by Eq.(\ref{transf}) along with the seed solution given by Eq. (\ref{seed}) into Eq. (\ref{BK}) and collect the coefficients of different powers of $\phi$. We start from the lowest powers of $\phi$. Collecting the coefficients of  $\phi^{-5}$:
\begin{equation}
24u_0 \phi_x^4+24u_0 \phi_x^3 \phi_y - 12 u_0^2 \phi_x^3-12 u_0^2 \phi_x^2 \phi_y = 0.
\end{equation}
Solving the above equation, we obtain
\begin{equation}
u_0 = 2 \phi_x.
\end{equation}
Collecting the coefficients of  $\phi^{-4}$, we get
\begin{subequations}
\begin{eqnarray}
-24u_{ox}\phi_x^3 - 36 u_0 \phi_x^2 \phi_{xx} - 18u_{0x}\phi_x^2\phi_y 
- 6 u_{0y}\phi_x^3-18u_0\phi_x^2\phi_{xy}
\nonumber \\
-18u_0\phi_x\phi_y\phi_{xx}+12u_0u_{0x}\phi_x^2+6u_0^2\phi_x\phi_{xx} 
+12u_0u_{0x}\phi_x^2+3u_0u_{0x}\phi_x\phi_y \nonumber \\+3u_0u_{0y}\phi_x^2+3u_0^2\phi_x\phi_{xy} 
+12u_0u_{0x}\phi_x\phi_y+3u_0^2\phi_{xx}\phi_y+6u_0u_{0y}\phi_x^2=0. \label{phi-4}
\end{eqnarray}  
\end{subequations}
Eq. (\ref{phi-4}) is an identity.

Next, collecting the coefficients of $\phi^{-3}$, we obtain
\begin{subequations}
\begin{eqnarray}
2u_0\phi_x\phi_t+2u_0\phi_y\phi_t+12u_{0xx}\phi_x^2+24u_{0x}\phi_x\phi_{xx}+6u_0\phi_{xx}^2+8u_0\phi_x\phi_{xxx}
\nonumber \\
-6u_{0xx}\phi_x\phi_y+6u_{0xy}\phi_x^2+12u_{0x}\phi_x\phi_{xy}+6u_{0x}\phi_y\phi_{xx}+2u_0\phi_y\phi_{xxx}+6u_{0y}\phi_x\phi_{xx}
\nonumber \\
+6u_0\phi_{xy}\phi_{xx}+6u_0\phi_x\phi_{xxy}-12u_{0x}^2\phi_x-6u_0u_{0x}\phi_{xx}-6u_0u_{0xx}\phi_x-3u_0u_{0xx}\phi_y
\nonumber \\
-6u_{0x}u_{0y}\phi_x-3u_0u_{0y}\phi_{xx}-3u_{0x}^2\phi_y
-3u_0u_{0xy}\phi_x-3u_{0x}u_{0y}\phi_x-3u_0u_{0x}\phi_{xy}
\nonumber \\
+2\alpha u_0\phi_x\phi_y+2\beta u_0\phi_x\phi_z+2 \beta u_0\phi_y\phi_z+2 \gamma_1 u_0 \phi_x^2 
+ 2 \gamma_2 u_0 \phi_y^2 = 0. \label{phi-3}
\end{eqnarray}  
\end{subequations}
Simplifying  Eq. (\ref{phi-3}) by substituting the value of $u_0$, we obtain
\begin{subequations}
\begin{eqnarray}
4 \phi_x \phi_t+4\phi_y\phi_t-12\phi_{xx}^2+12\phi_{xxy}\phi_x+4\alpha \phi_x\phi_y+4 \beta \phi_x \phi_z +4 \beta \phi_y \phi_z
\nonumber \\
+4 \gamma_1 \phi_x^2 +4 \gamma_2 \phi_y^2 + 16 \phi_x \phi_{xxx}-12 \phi_{xx}\phi_{xy}+4
\phi_x\phi_y\phi_{xxx}=0. \label{R1}
\end{eqnarray}  
\end{subequations}

Next, collecting the coefficients of $\phi^{-2}$, we get
\begin{subequations}
\begin{eqnarray}
-u_{0x}\phi_t-u_{0t}\phi_x-u_0\phi_{xt}-u_{0y}\phi_t-u_{0t}\phi_y-u_0\phi_{yt}-4u_{0xxx}\phi_x-6u_{0xx}\phi_{xx}
\nonumber \\
-4u_{0x}\phi_{xxx}-u_0\phi_{xxxx}-u_{0xxx}\phi_y-3u_{0xxy}\phi_x-3u_{0xx}\phi_{xy}-3u_{0xy}\phi_{xx}
\nonumber \\
-3u_{0x}\phi_{xxy}-u_{0y}\phi_{xxx}-u_0\phi_{xxxy}+6u_{0x}u_{0xx}+3u_{0x}u_{0xy}+3u_{0xx}u_{0y}
\nonumber \\
-\alpha u_{0x}\phi_y - \alpha u_{0y} \phi_x -\alpha u_0\phi_{xy}-\beta u_{0x}\phi_z - \beta u_{0z}\phi_x-\beta u_0\phi_{xz}
\nonumber \\
-\beta u_{0z} \phi_y - \beta u_{0y} \phi_z - \beta u_0 \phi_{yz} -2 \gamma_1 u_{0x} \phi_x - \gamma_1 u_0 \phi_{xx} 
\nonumber \\
- 2 \gamma_2 u_{0y} \phi_y - \gamma_2 u_0 \phi_{yy}=0. \label{phi-2}
\end{eqnarray}  
\end{subequations}
Simplifying the above Eq. (\ref{phi-2}) by substituting the value of $u_0$ again, we obtain
\begin{subequations}
\begin{eqnarray}
-2\phi_{xx}\phi_t-4\phi_x\phi_{xt}-2\phi_{xy}\phi_t-2\phi_{xt}\phi_y-2\phi_x\phi_{yt}-10\phi_{xxxx}\phi_x
\nonumber \\
-2\phi_{xxxx}\phi_y-8\phi_{xxxy}\phi_x+4\phi_{xx}\phi_{xxx}+4\phi_{xxx}\phi_{xy}-2 \alpha \phi_{xx}\phi_y
\nonumber \\
-4 \alpha \phi_x\phi_{xy} -2 \beta \phi_{xx}\phi_z-4 \beta \phi_x \phi_{xz}-2 \beta \phi_{xz}\phi_y -2 \beta \phi_{xy} \phi_z 
\nonumber \\
- 2 \beta \phi_x \phi_{yz} - 6 \gamma_1 \phi_{xx} \phi_x -4 \gamma_2 \phi_{xy} \phi_y - 2 \gamma_2 \phi_x \phi_{yy} =0.
\end{eqnarray}  \label{R2}
\end{subequations}
Finally, collecting the coefficients of $\phi^{-1}$, we have
\begin{equation}
u_{0xt}+u_{0yt}+u_{0xxxx}+u_{0xxxy}+\alpha u_{0xy}+\beta u_{0xz} 
+ \beta u_{0yz}+\gamma_1 u_{0xx} + \gamma_2 u_{oyy} = 0. \label{phi-1}
\end{equation}
Simplifying the above Eq. (\ref{phi-1}), we obtain
\begin{subequations}
\begin{eqnarray}
2 \phi_{xxt} + 2 \phi_{xyt} + 2 \phi_{xxxxx} + 2 \phi_{xxxxy} +2 \alpha \phi_{xxy} + 2 \beta \phi_{xxz} \nonumber \\
+ 2 \beta \phi_{xyz} + 2 \gamma_1 \phi_{xxx} + 2 \gamma_2 \phi_{xyy} = 0. \label{R3}
\end{eqnarray}  
\end{subequations}
Now, to look for a solution which satisfies Eqs. (\ref{R1}), (\ref{R2}) and (\ref{R3}), we have
\begin{equation}
\phi = x+\xi(y), 
\end{equation}
for the parametric choice $\alpha$=$\gamma_1$=$\gamma_2$=0.

Using the above form of $\phi$, we can write down the general solution of (3+1) dimensional Bogoyavlensky-Konopelchenko equation in the form of the theorem given by  (\ref{sol}). Hence, the theorem.

\section{Solutions of (3+1) dimensional Bogoyavlensky-Konopelchenko Equation
}
\subsection{Periodic solutions}
\begin{figure}[h!]
\begin{center}
\includegraphics[width=0.42\linewidth]{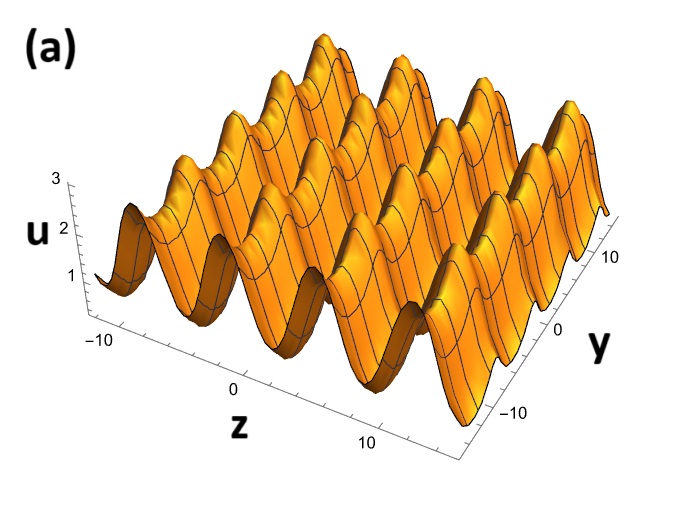}
\includegraphics[width=0.4\linewidth]{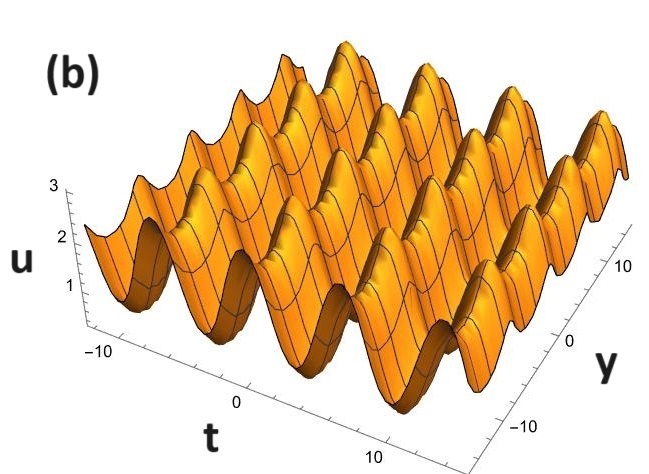}
\includegraphics[width=0.4\linewidth]{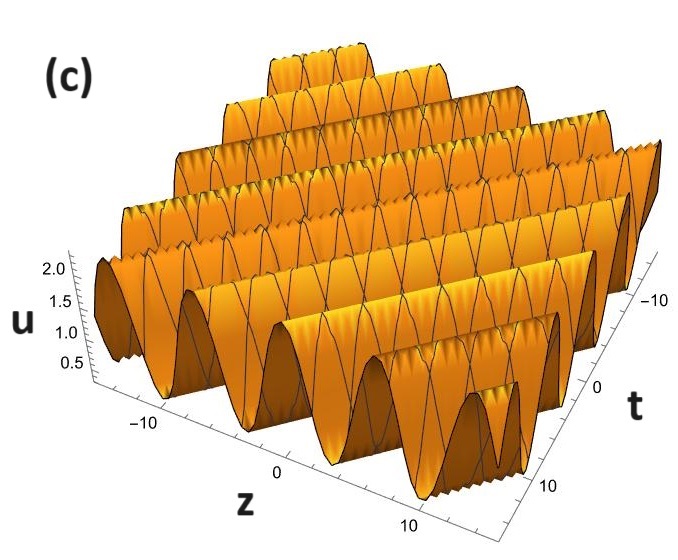}
\includegraphics[width=0.32\linewidth]{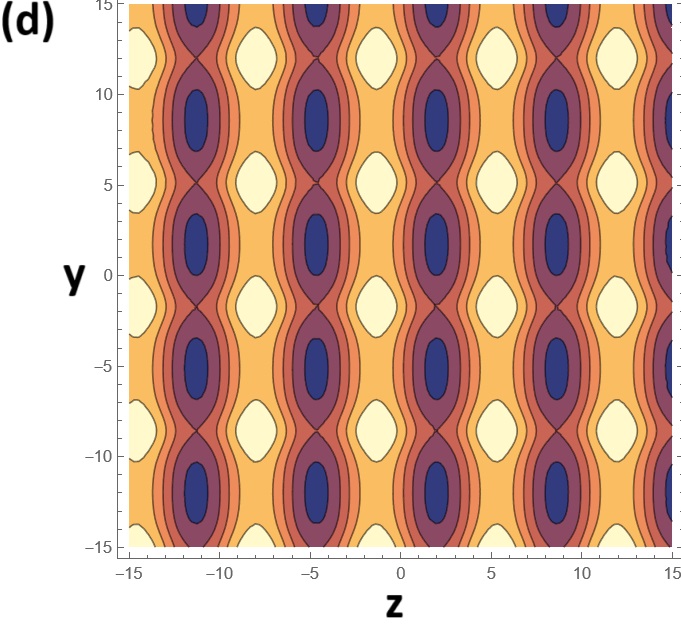}
\caption{Periodic solution (a) in yz plane for t=-5 (b) in yt plane for z=-5 (c) in zt plane for y=-5 (d) contour plot corresponding to Fig.\ref{per}a }
\label{per}
\end{center}
\end{figure}

To construct periodic solutions using Eq.(\ref{sol}), we choose the arbitrary functions $u_1$ and $\xi$ to be Jacobian elliptic funtions, namely sn function.
\begin{equation}
    u_1=1+\mbox{sn}(z+t,m_1), \label{periodic1}
\end{equation}
\begin{equation}
    \xi=1+\mbox{sn}(y,m_2), \label{periodic2}
\end{equation}
where $m_1$ and $m_2$ are moduli of the respective Jacobian elliptic function. Substituting the above choice of functions in Eq.(\ref{sol}), we obtain
\begin{equation}
   u= 1 + \frac{2}{(1 + x + \mbox{sn}(y, m_2))} + \mbox{sn}(t + z, m_1).
\end{equation}
For the parametric choice $m_1=0.2;m_2=0.3;x=1$, we obtain periodic solutions identified in several two dimensional planes as shown in Fig.(\ref{per}). Note that as time evolves, this periodic wave travels along the z-direction.

\subsection{Kink Solution}
To construct kink solution using Eq.(\ref{sol}), we choose
\begin{equation}
u_1=\frac{(1 + \mbox{tanh}(z - t))}{(1 + \mbox{sech}(z - t)^2)},
\end{equation}
\begin{equation}
\xi=\frac{1 + \mbox{tanh}y}{1 + \mbox{sech}y^2}.
\end{equation}

\begin{figure}[h!]
\begin{center}
\includegraphics[width=0.4\linewidth]{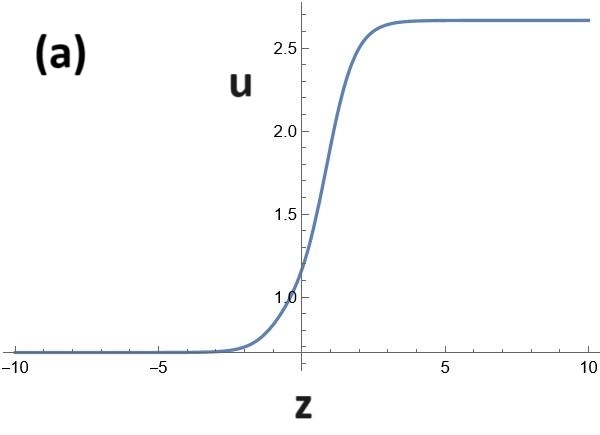}
\includegraphics[width=0.4\linewidth]{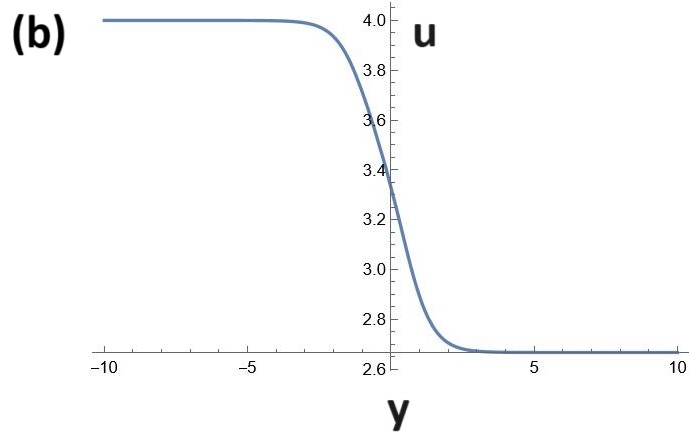}
\includegraphics[width=0.5\linewidth]{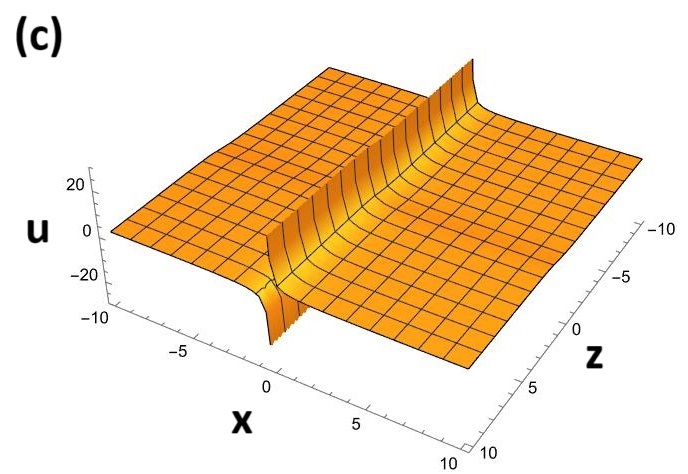}
\caption{Kink solution at time t=0 (a) along z direction, (b) along y-direction (c) in xz plane for y=0, t=0}
\label{kink}
\end{center}
\end{figure}

We now obtain the kink solution as 
\begin{equation}
u =\frac{2(1 + \mbox{sech}y^2)}{x(1 + \mbox{sech}y^2) + (1 + \mbox{tanh}y)} + \frac{1 + \mbox{tanh}(z-t)}{
 1 + \mbox{sech}(z-t)^2}.
\end{equation}
For the choice  $x=1,y=5$,  the kink solution along the z and y direction shown in Figs.\ref{kink}(a,b) while the profile of the kink solution in xz plane is shown in Fig.\ref{kink}(c).

\subsection{Linear Rogue Wave}
\begin{figure}[h!]
\begin{center}
\includegraphics[width=0.4\linewidth]{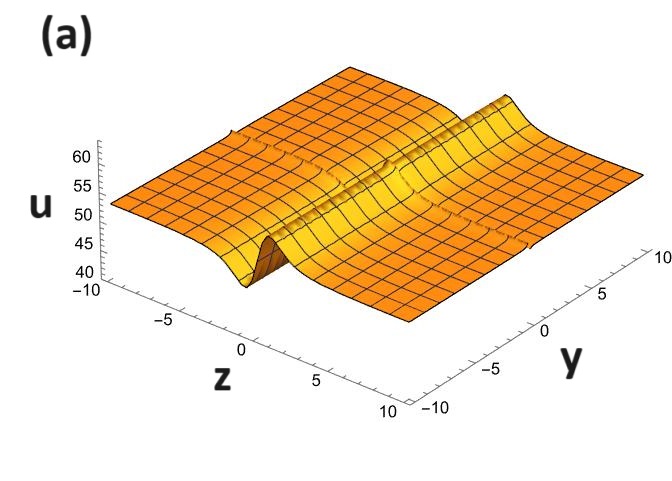}
\includegraphics[width=0.4\linewidth]{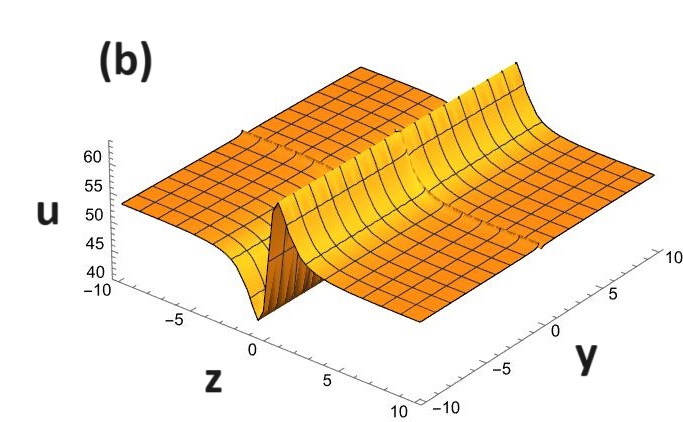}
\includegraphics[width=0.4\linewidth]{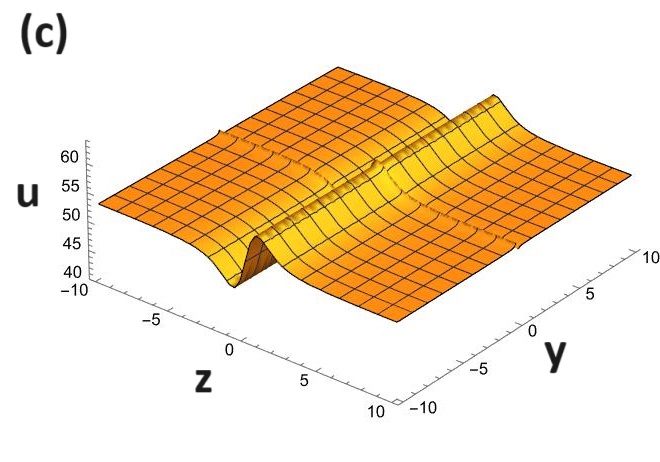}
\includegraphics[width=0.4\linewidth]{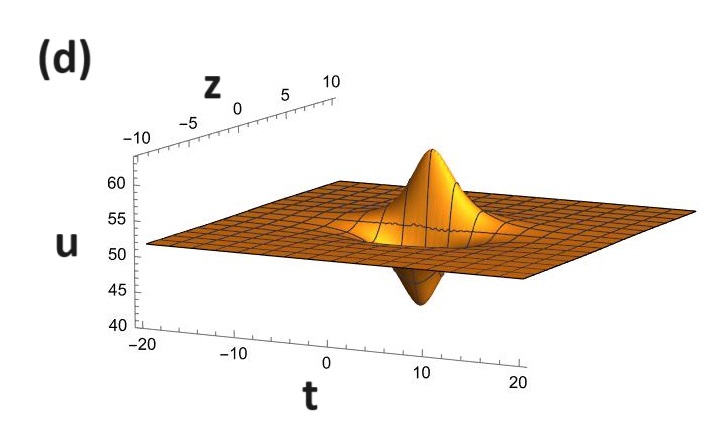} \label{rogue-zt}
\includegraphics[width=0.35\linewidth]{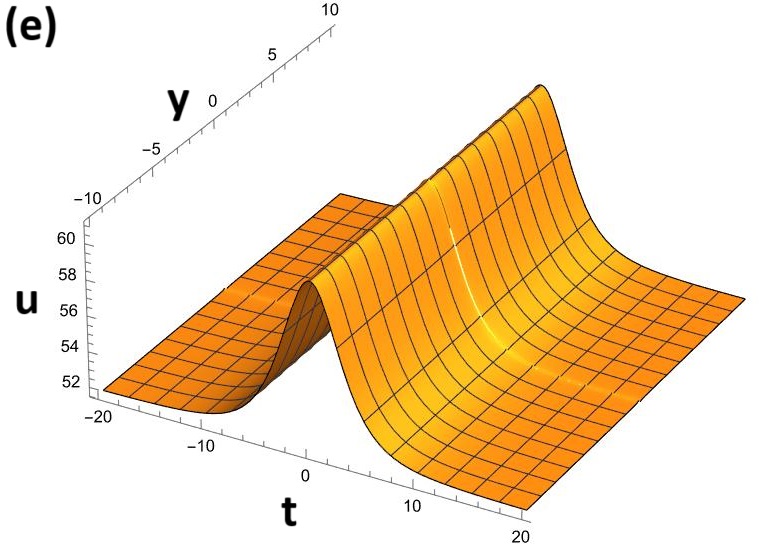}
\caption{Linear rogue wave solution for time (a) t=-3, (b) t=0 and (c) t=3; (d) Linear rogue wave solution in the zt plane for y=50 (e) yt plane for z=1 }
\label{rogue}
\end{center}
\end{figure}

\begin{figure}[h!]
\begin{center}
\includegraphics[width=0.3\linewidth]{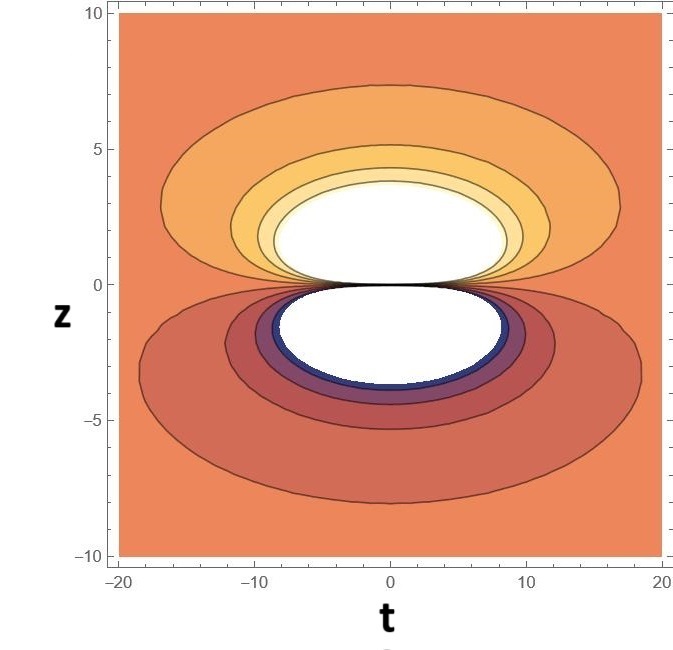}
\caption{Contour Plot of linear rogue wave solution for the panel Fig.3(d) }
\label{rogue_cont}
\end{center}
\end{figure}

To construct a linear rogue wave using Eq.(\ref{sol}), we choose
\begin{equation}
u_1=\frac{c z}{(1 + a_1 z^2 + d t^2)^2} + \frac{1}{(2 d)^2},
\end{equation}
\begin{equation}
\xi=\frac{1}{((y - b)^2 g + kny)}
\end{equation}
and substitute in Eq.(\ref{sol}) to  obtain
\begin{equation}
  u=\frac{1}{4 d^2} + \frac{2(k n y + g (-b + y)^2)}{1+x (k n y + g (-b + y)^2))} + \frac{c z}
 {(1 + d t^2 + a_1 z^2)^2}.
\end{equation}
For the parametric choice $c = 30; a_1 = 0.8; d = 0.07; b = 0.9; k = 10; n = 4; g = 0.6; x = 1$, we obtain a linear rogue wave as shown in Figs.(\ref{rogue}a-\ref{rogue}c).
 From Figs.(\ref{rogue}a-\ref{rogue}c), it is obvious that the amplitude of the linear rogue wave which resembles a  composite dark-bright soliton  is unstable while the one shown in Fig(\ref{rogue}e) resembles a line soliton.The contour plot of Fig.(\ref{rogue}d) where one sees the sudden eruption of the rogue wave at $t=0$ is shown in Fig.(\ref{rogue_cont}).  

\subsection{Line lumps}
To construct a line lump solution using Eq.(\ref{sol}), we choose
\begin{equation}
u_1=\frac{1}{(1 + (z - t - k_1)^2)},
\end{equation}
\begin{equation}
\xi=c_1 + (ky - h)^2
\end{equation}
and substitute in Eq.(\ref{sol}) to obtain
\begin{equation}
    u= \frac{2}{(c_1 + x + (-h + k y)^2)} + \frac{1}{(1 + (-k_1 - t + z)^2)}.
\end{equation}

\begin{figure}[h!]
\begin{center}
\includegraphics[width=0.4\linewidth]{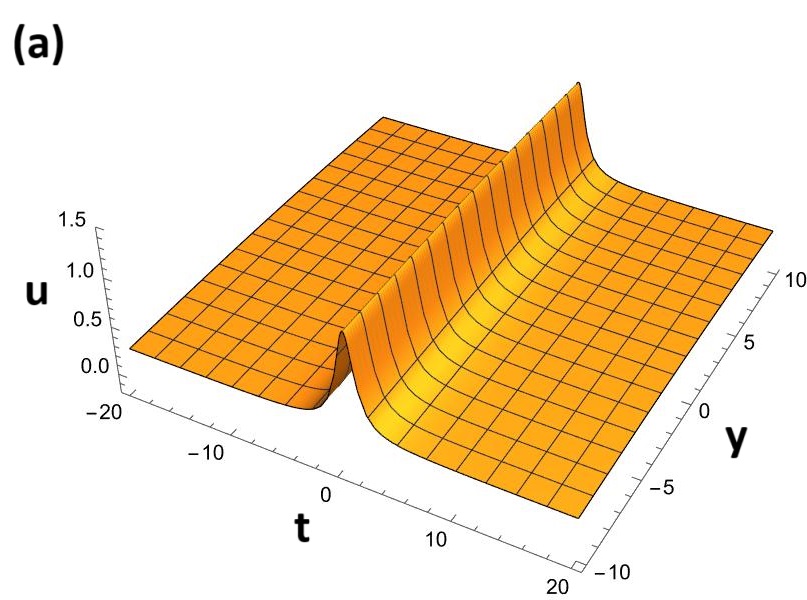}
\includegraphics[width=0.4\linewidth]{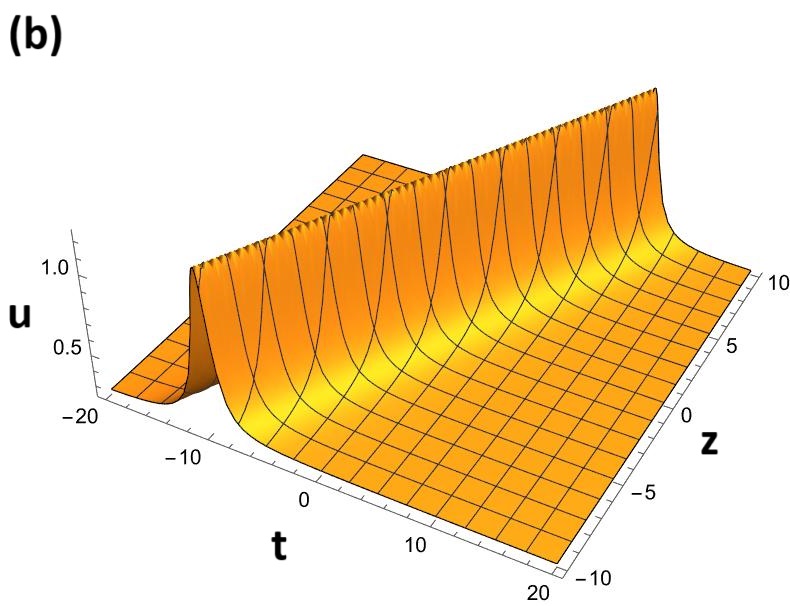}
\includegraphics[width=0.4\linewidth]{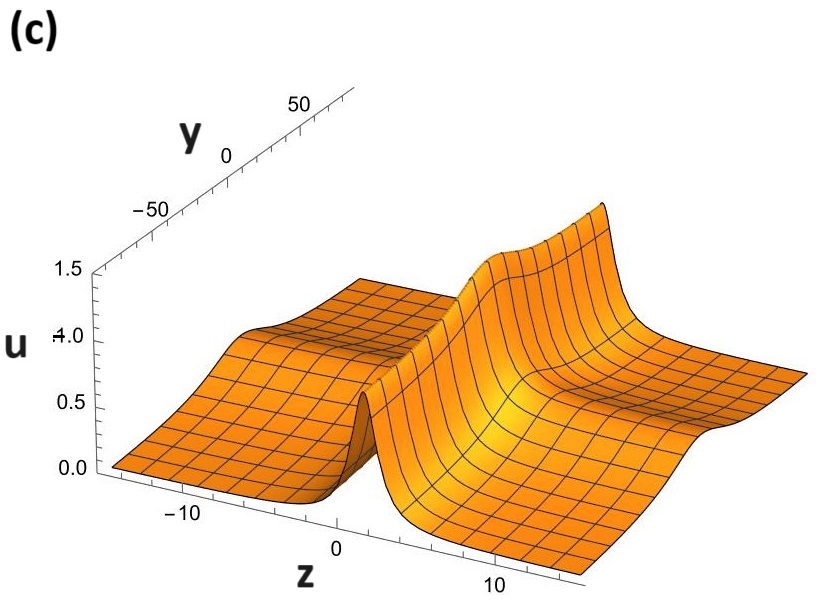}
\includegraphics[width=0.4\linewidth]{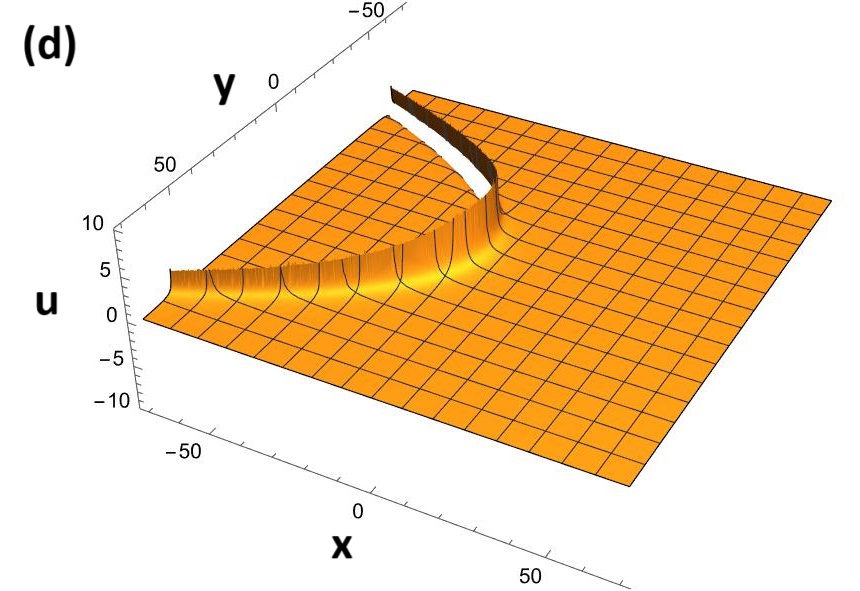}
\includegraphics[width=0.5\linewidth]{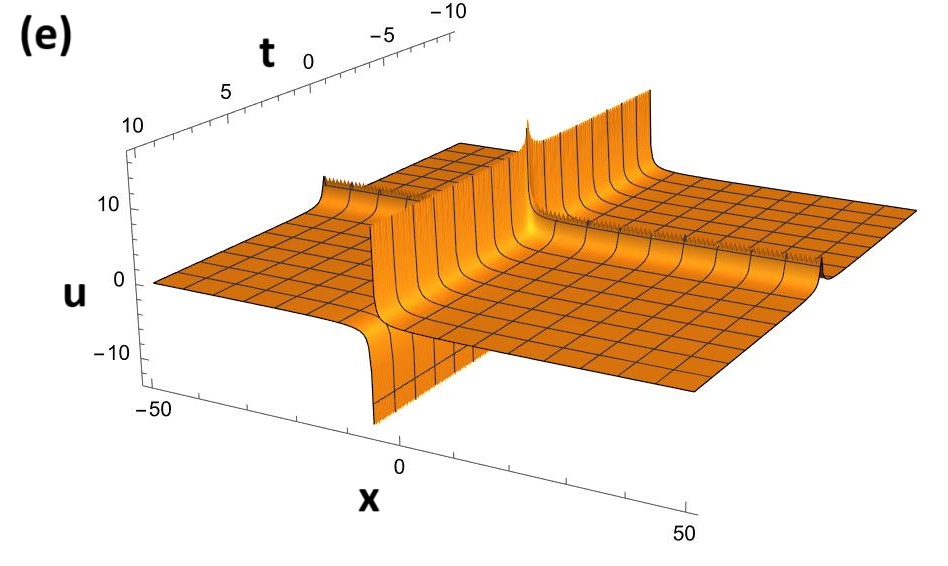}
\caption{ Line Lump solution (a) in yt plane for z=1 (b) in zt plane for y=1 and (c) in zy plane for t=1 (d) in xy plane for z=10 (e) in xt plane for y=0 and z=0 }
\label{olump}
\end{center}
\end{figure}

\begin{figure}[h!]
\begin{center}
\includegraphics[width=0.4\linewidth]{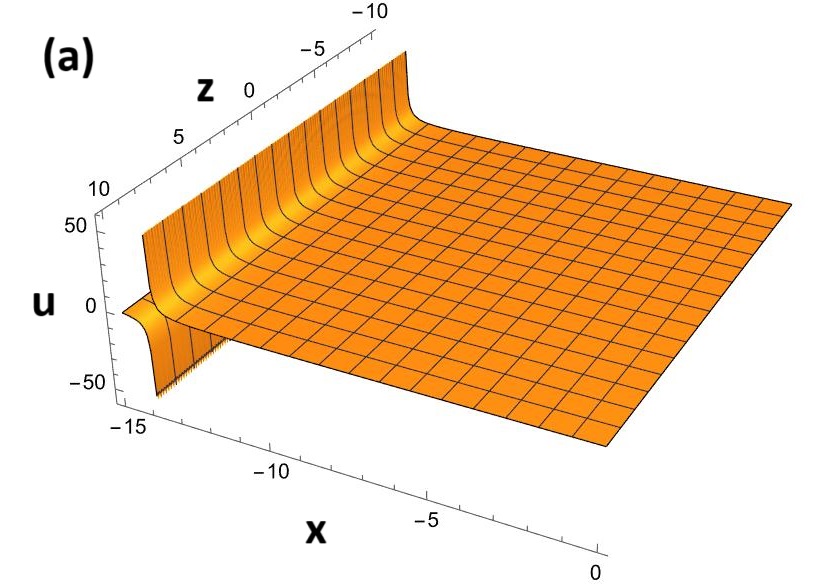}
\includegraphics[width=0.4\linewidth]{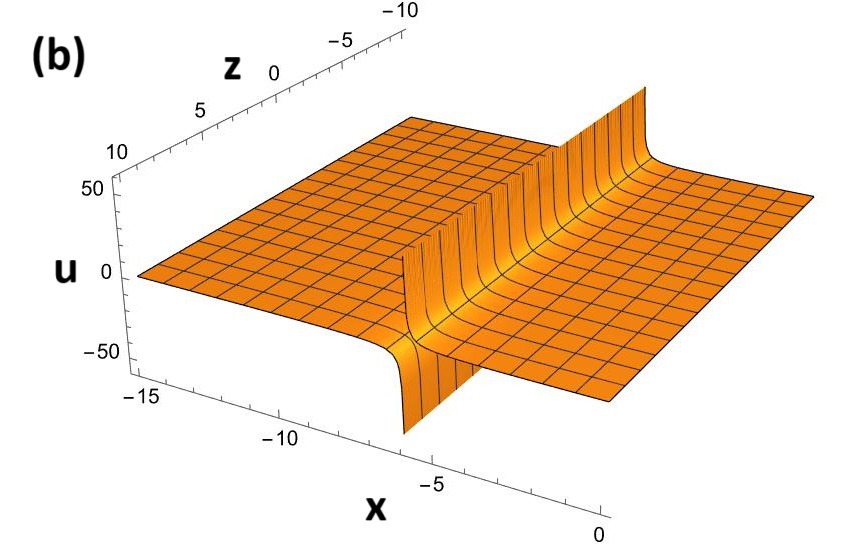}
\includegraphics[width=0.4\linewidth]{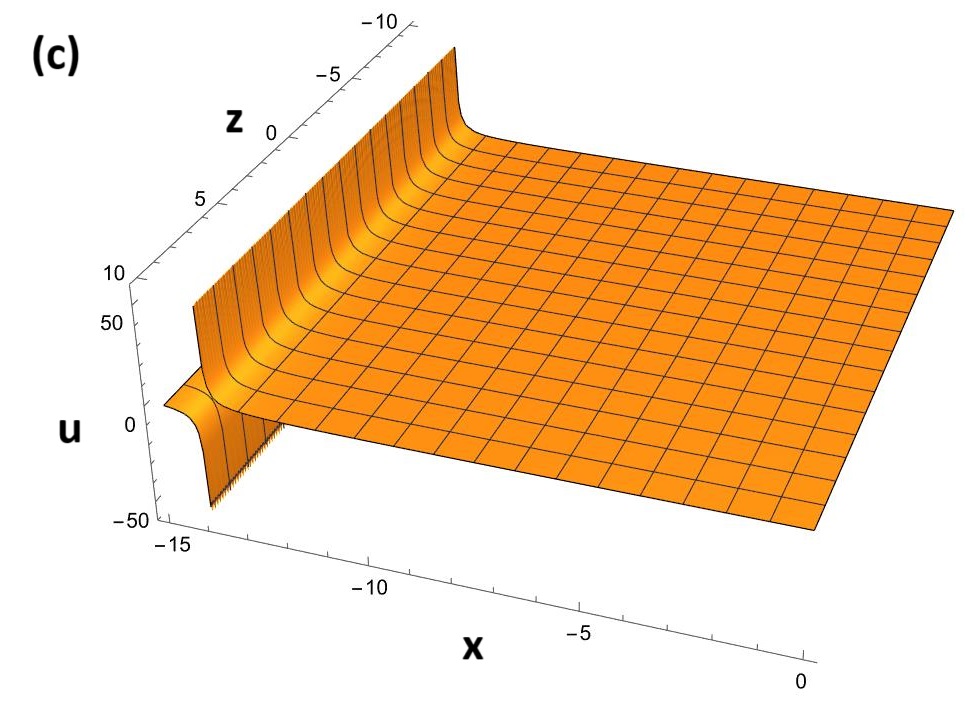}
\caption{ Dipole Lump solution in the xz plane for t=-10 at (a) y=-20 (b) y=0 and (c) y=20}
\label{olump-xz}
\end{center}
\end{figure}

For the parametric choice $c_1=5;x = 1; k = 0.1; k_1 = 0.4; h = 0.1;$, we obtain a line  lump solution as shown in Fig.(\ref{olump}). It is again interesting to observe that  the solutions shown in Figs.\ref{olump}(a) and \ref{olump}(b) resemble a single hump line soliton while the one shown in Fig(\ref{olump})c looks like a double hump line soliton.  The line lump solution  shown in xy plane (Fig.\ref{olump}(d) ) is curved and while the one shown in xt plane 
(Fig.\ref{olump}(e)) comprises of a kink and a line soliton . 
One can also obtain a dipole lump solution with positive and negative density in the xz plane at different positions of y as shown in Fig.\ref{olump-xz}.

\subsection{Two line  lumps}
\begin{figure}[h!]
\begin{center}
\includegraphics[width=0.4\linewidth]{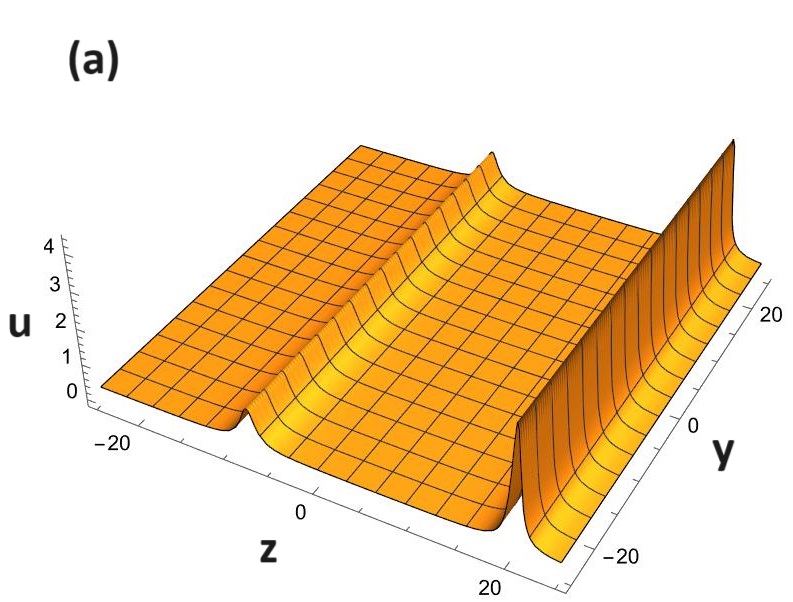}
\includegraphics[width=0.4\linewidth]{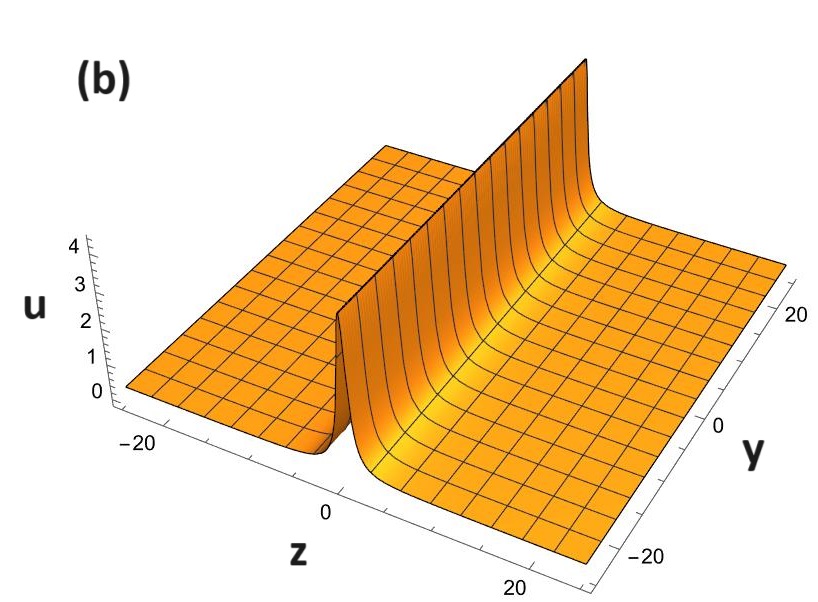}
\includegraphics[width=0.4\linewidth]{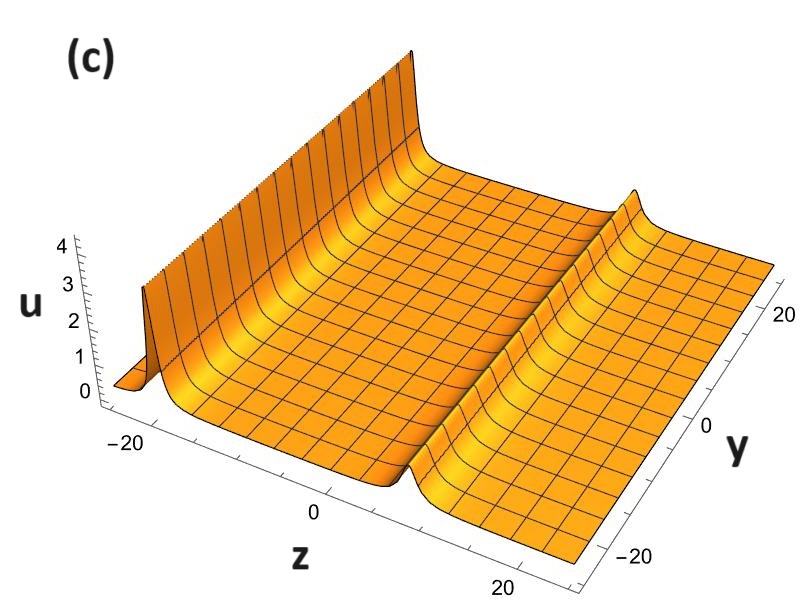}
\caption{Time evolution of two line lump solution for time (a) t=-4, (b) t=0 and (c) t=4}
\label{2lump}
\end{center}
\end{figure}

\begin{figure}[h!]
\begin{center}
\includegraphics[width=0.3\linewidth]{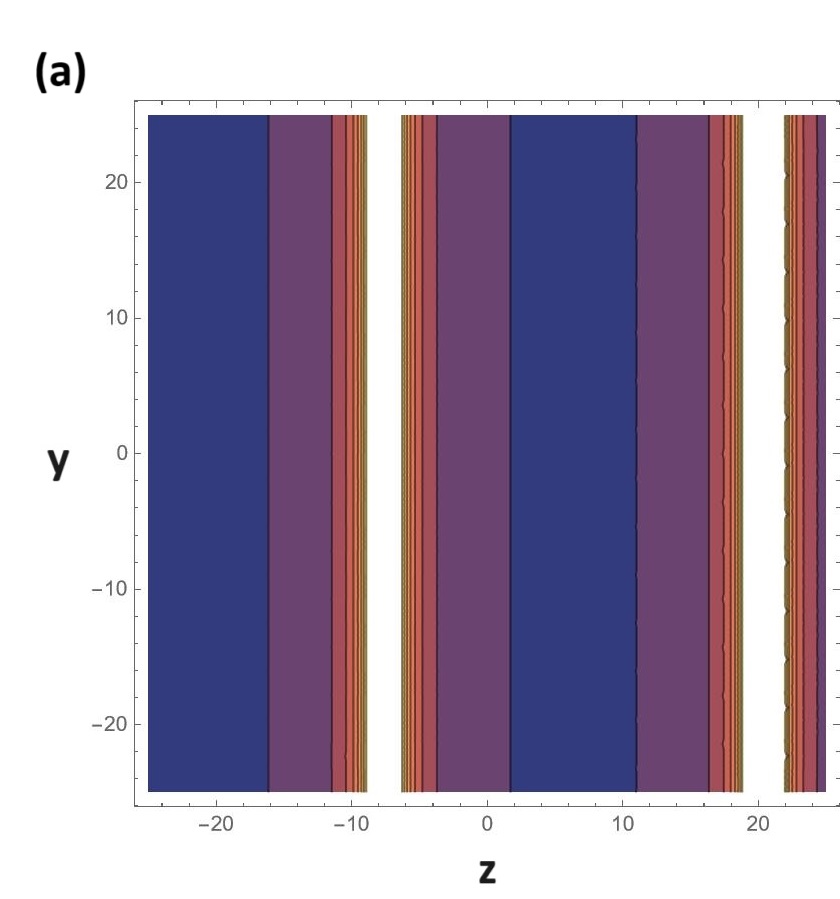}
\includegraphics[width=0.3\linewidth]{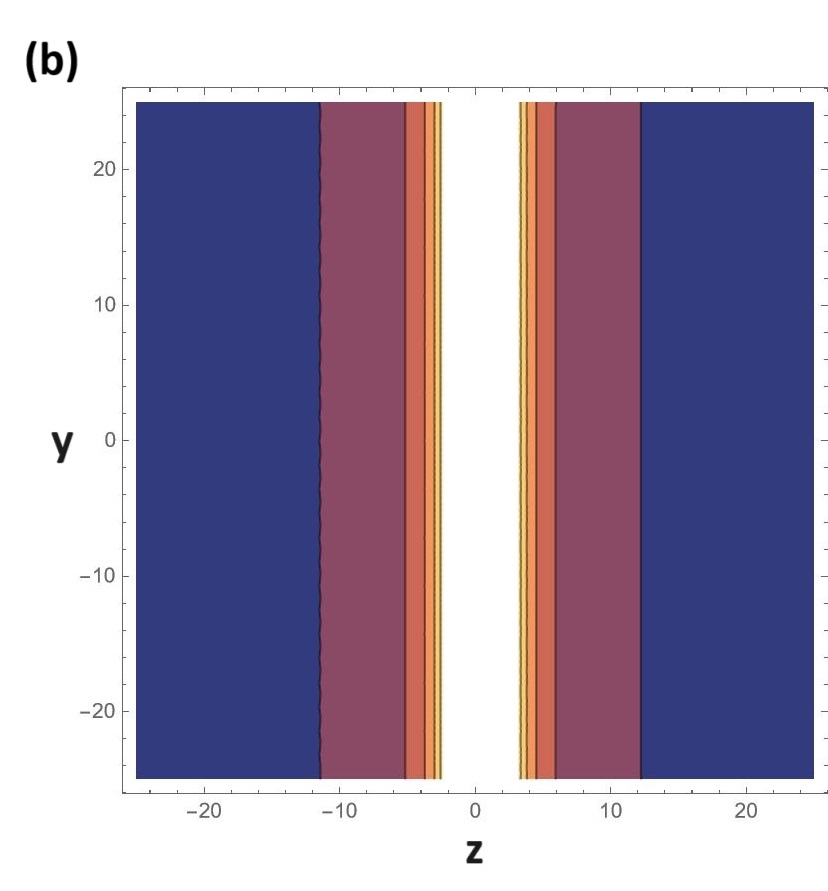}
\includegraphics[width=0.3\linewidth]{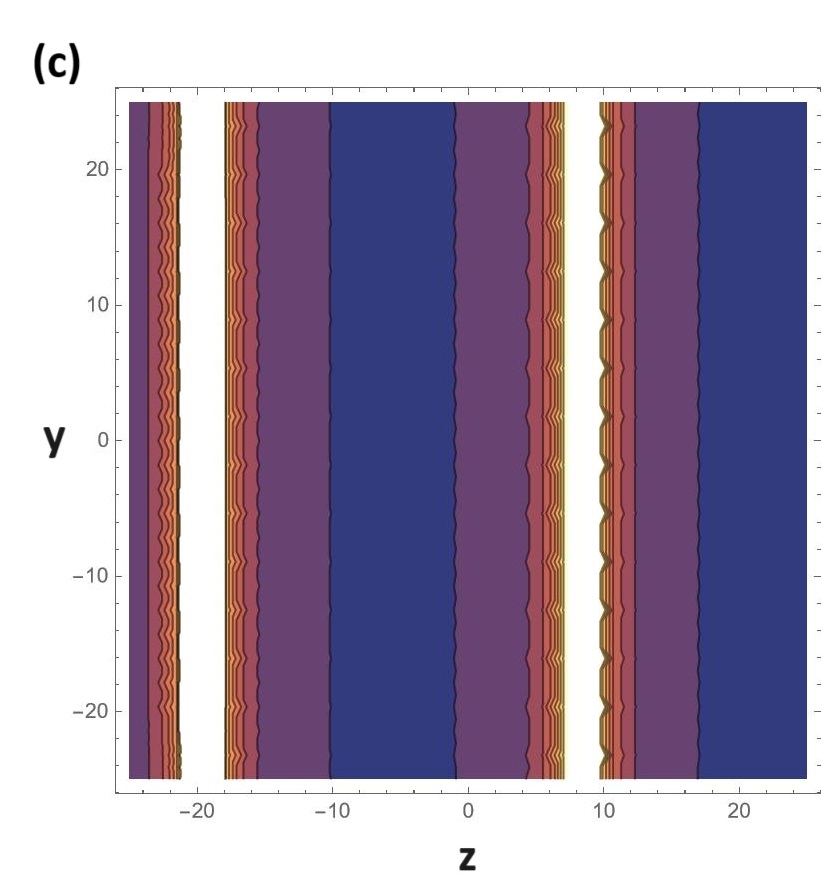}
\caption{Contour plot of the time evolution of two line lump solution for time (a) t=-4, (b) t=0 and (c) t=4}
\label{2lump-cont}
\end{center}
\end{figure}

To construct two line lumps using Eq.(\ref{sol}), we choose
\begin{equation}
u_1=\frac{1}{(q_1 + (z + m_1 t + k_1)^2)}+\frac{1}{(q_2 + (z + m_2 t + k_2)^2)},
\end{equation}
\begin{equation}
\xi=c_1 + (ky - h)^2
\end{equation}
and substitute in Eq.(\ref{sol}) to  give
\begin{equation}
    u= \frac{2}{(c_1 + x + (-h + k y)^2)} + \frac{1}{p_1}+\frac{1}{p_2}.\label{2lumpSol}
\end{equation}
where $p_1=q_1 + (z + m_1 t + k_1)^2$ and $p_2=q_2 + (z + m_2 t + k_2)^2$. The time evolution of two line lumps for the parametric choice $x = 1; k = 0; k_1 = k_2 = 0.4; m_1=-2;m_2=5; h = 0; q_1=1; q_2=0.3$ is shown in Fig.\ref{2lump}.
The corresponding contour plot is shown in Fig.\ref{2lump-cont}. From Fig.\ref{2lump}, it is obvious that the two line lumps which again resemble a line soliton are moving in opposite directions along $z$  axis and undergo elastic collision which is again reinforced by the contour plot shown in Fig.\ref{2lump-cont}. 

\subsection{Asymptotic analysis of two line lumps}
\begin{figure}[h!]
\begin{center}
\includegraphics[width=0.4\linewidth]{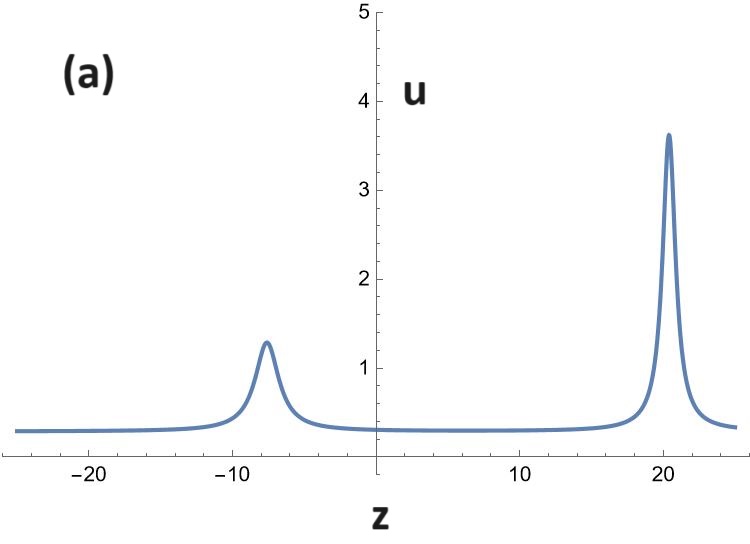}
\includegraphics[width=0.4\linewidth]{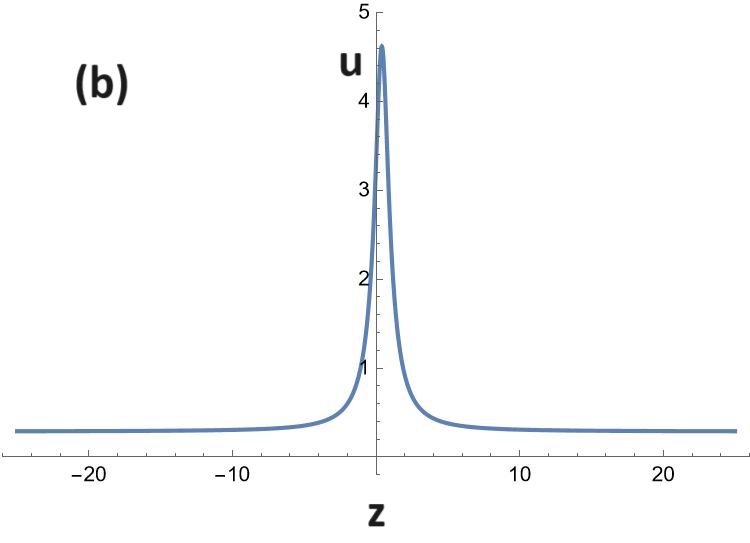}
\includegraphics[width=0.4\linewidth]{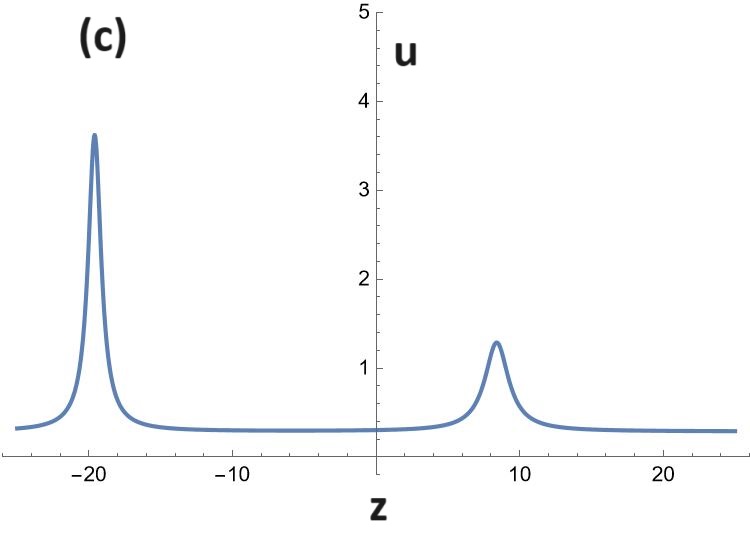}
\caption{Time evolution of two lump solution for time (a) t=-4, (b) t=0 and (c) t=4}
\label{2lump-z}
\end{center}
\end{figure}

As the line lumps are travelling in  opposite directions along the z-axis.  we analyse the evolution of line lumps for the limit $x=1$ and $y=0$. The time evolution of line lumps for this limit is shown in Fig.\ref{2lump-z}
Similar analysis holds good
for any other value of $x$ and $y$.  We analyze the limits
$t\rightarrow - \infty$  and $t\rightarrow + \infty$ separately so as to understand the interaction of line lumps
centered around $p_1 \approx 0$ or $p_2 \approx 0$. Without loss of generality, let us assume
$m_1 > m_2$. Then, we find in the limit t $\rightarrow$ $\pm$ $\infty$, $p_1$ and $p_2$ take the
following limiting values.
\newline
(1) As t $\rightarrow$ - $\infty$ :\\
$p_1 \approx 0$, $p_2 \rightarrow + \infty$,\\
$p_2 \approx 0, p_1 \rightarrow - \infty$. \\
\newline
(2) As t $\rightarrow$ + $\infty$ : \\
$p_1 \approx 0$, $p_2 \rightarrow - \infty$,\\
$p_2 \approx 0, p_1 \rightarrow + \infty$.\\

\paragraph
1. Before interaction (as $t \rightarrow$ -$\infty$):
For $p_1 \approx 0$, $p_2 \rightarrow +\infty$ and $t = -2; k = 0; k_1 = 0.4;k_2 = 0.4; h = 0$, the two line lump
solution (\ref{2lumpSol}) becomes 
\begin{equation}
	 u= \frac{2}{(c_1 + x + (-h + k y)^2)} + \frac{1}{1 + (z + m_1 t + k_1)^2}.
\end{equation}
For $p_2 \approx 0$, $p_1 \rightarrow  -\infty$ , the solution (\ref{2lumpSol}) becomes 
\begin{equation}
	 u= \frac{2}{(c_1 + x + (-h + k y)^2)} + \frac{1}{1 + (z + m_2 t + k_2)^2}.
\end{equation}

\paragraph
2. After interaction (as $t \rightarrow$ +$\infty$):
For $p_1 \approx 0$, $p_2 \rightarrow -\infty$ ,the solution (\ref{2lumpSol}) 
\begin{equation}
	u= \frac{2}{(c_1 + x + (-h + k y)^2)} + \frac{1}{1 + (z + m_1 t + k_1)^2}.
\end{equation}
For $p_2 \approx 0$, $p_1 \rightarrow -\infty$, the solution (\ref{2lumpSol}) becomes 
\begin{equation}
	u= \frac{2}{(c_1 + x + (-h + k y)^2)} + \frac{1}{1 + (z + m_2 t + k_2)^2}.
\end{equation}
From the above analysis, we observe that the amplitude of line lumps before and after interaction are the same. This implies that due to interaction, there is no change in shape of the line lumps showing that  they undergo elastic collision.

\subsection{N- line lump solution}
To construct N- line lumps using Eq.(\ref{sol}), we choose
\begin{equation}
u_1=\sum_{i=1}^{N}\frac{1}{(q_i + (a_i z + m_i t + k_i)^2)},
\end{equation}
\begin{equation}
\xi=c_1 + (ky - h)^2
\end{equation}
and substitute in Eq.(\ref{sol}) to  obtain
\begin{equation}
    u= \frac{2}{(c_1 + x + (-h + k y)^2)} + \sum_{i=1}^{N}\frac{1}{(q_i + (a_i z + m_i t + k_i)^2)}, \label{nlump}
\end{equation}
where $c_1, h, k, q_i, a_i, m_i, k_i$ are arbitrary parameters. Eq.(\ref{nlump}) gives the expression for  N- line lumps which travels along the z-direction.

\begin{figure}[h!]
\begin{center}
\includegraphics[width=0.3\linewidth]{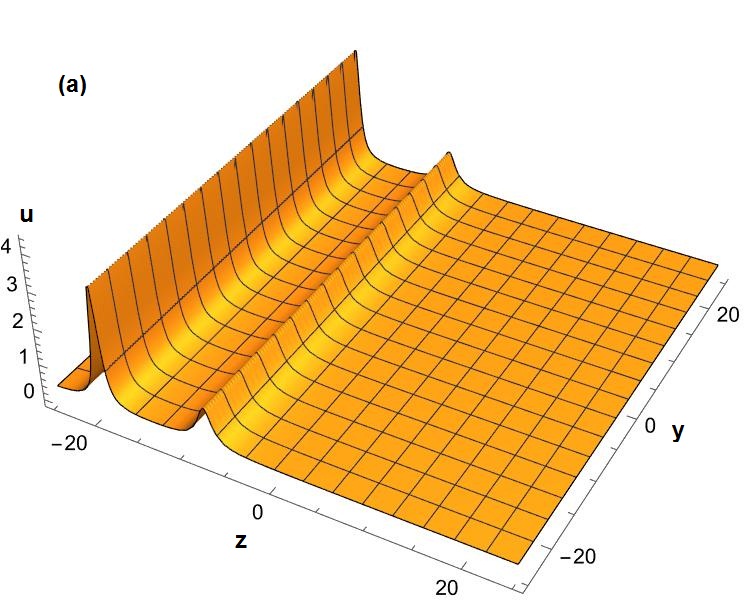}
\includegraphics[width=0.3\linewidth]{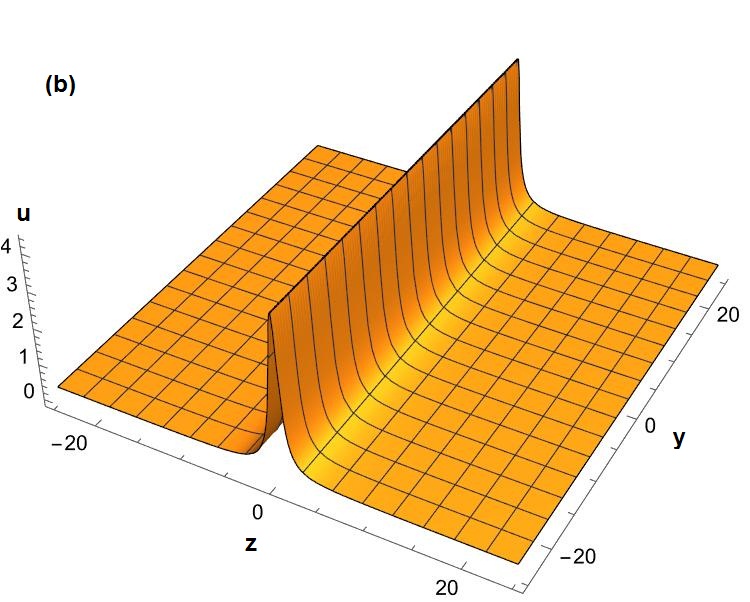}
\includegraphics[width=0.3\linewidth]{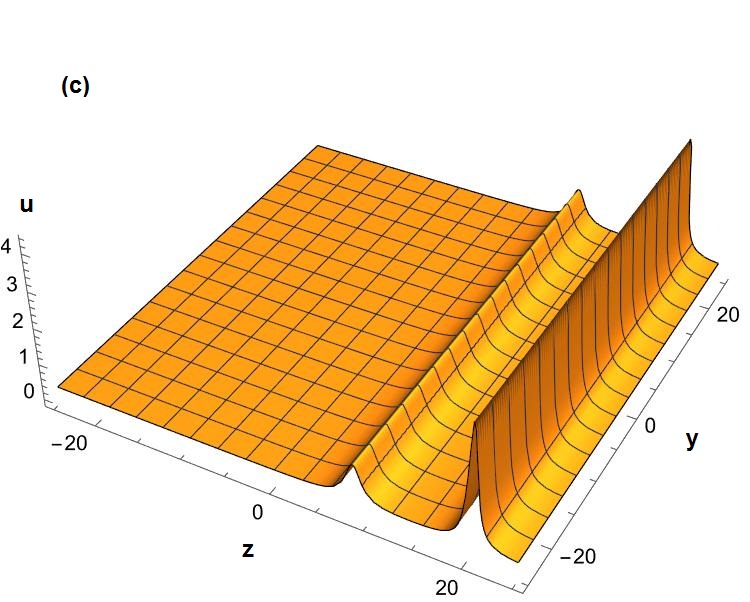}
\caption{Time evolution of two line lump solution along the same direction for time (a) t=-4, (b) t=0 and (c) t=4}
\label{2lump-same dir}
\end{center}
\end{figure}

\begin{figure}[h!]
\begin{center}
\includegraphics[width=0.3\linewidth]{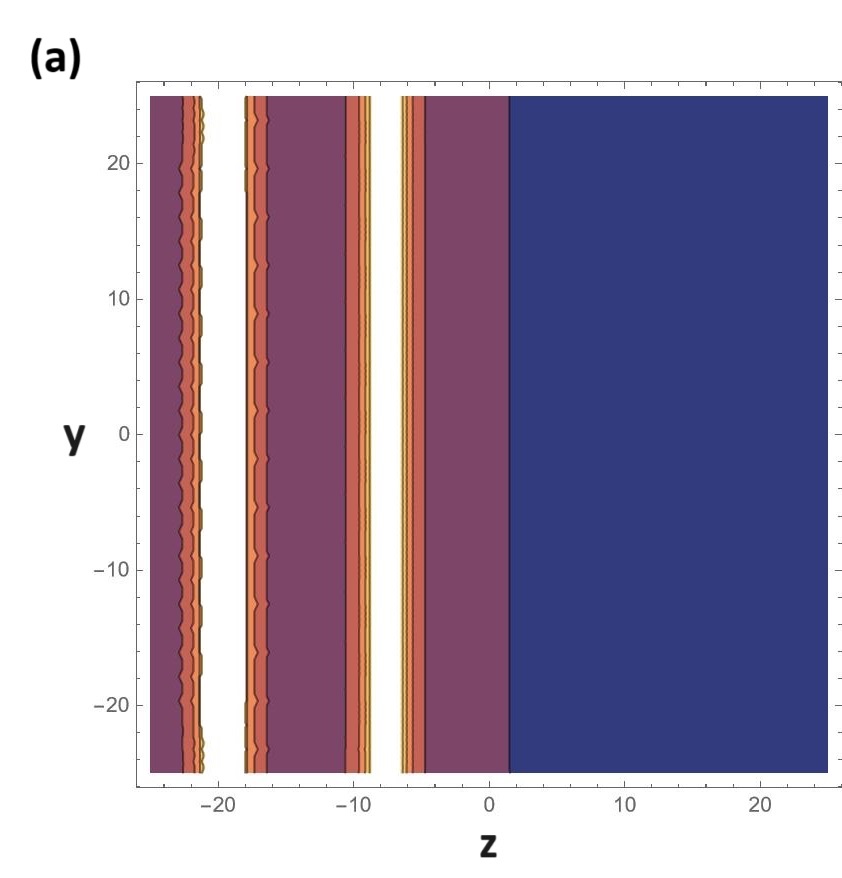}
\includegraphics[width=0.3\linewidth]{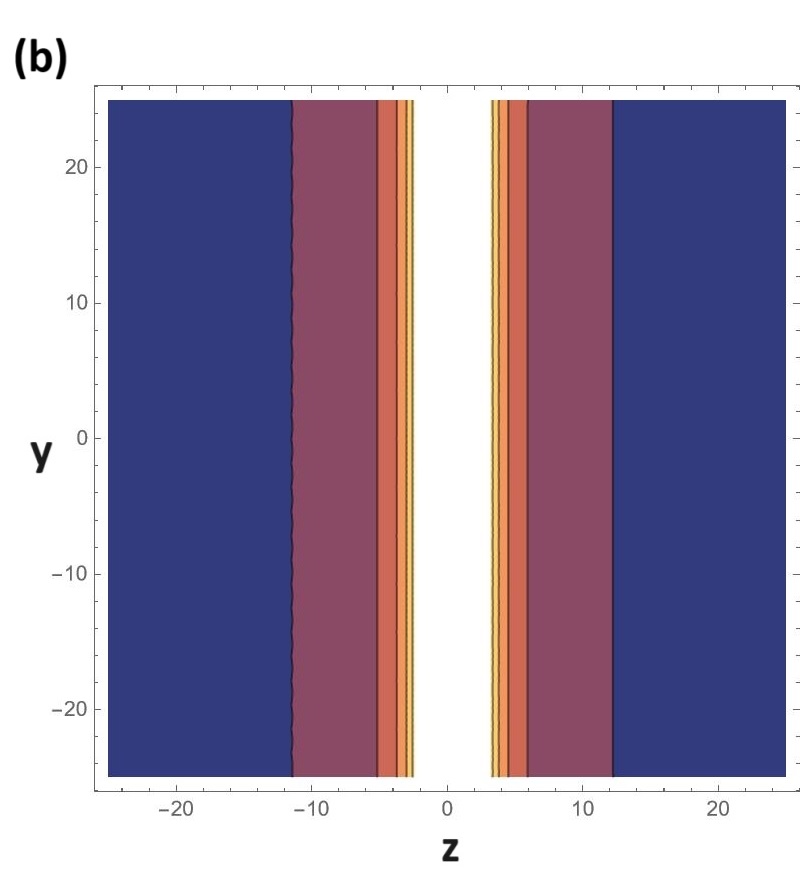}
\includegraphics[width=0.3\linewidth]{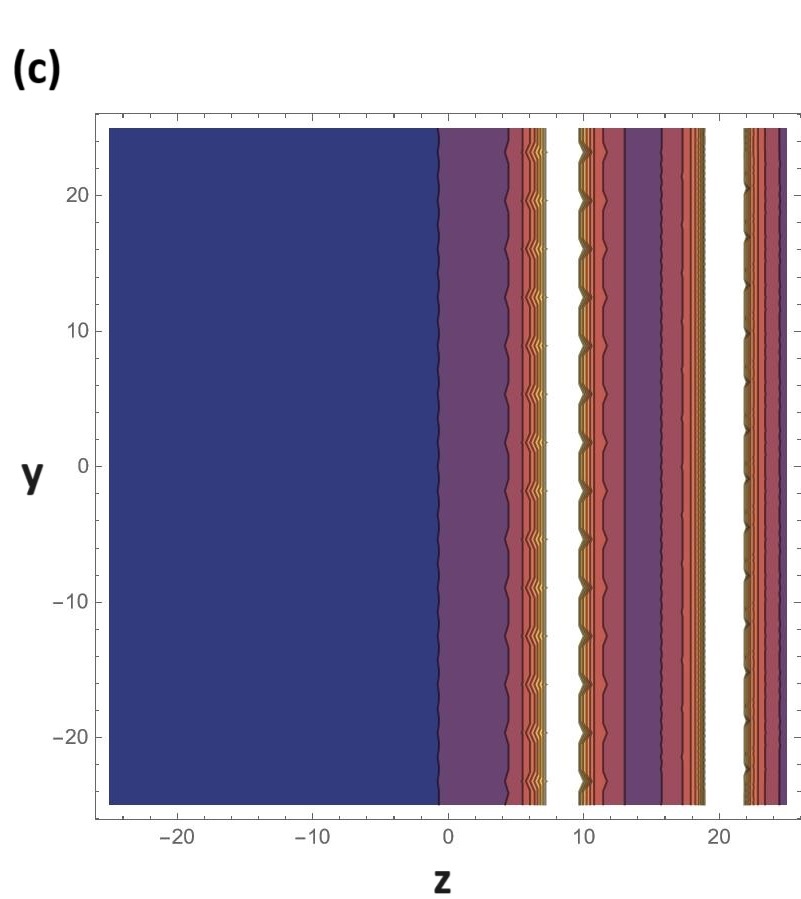}
\caption{Contour plot of the time evolution of two line lump solution along the same direction for time (a) t=-4, (b) t=0 and (c) t=4}
\label{2lump-same dir-cont}
\end{center}
\end{figure}

\begin{figure}[h!]
\begin{center}
\includegraphics[width=0.4\linewidth]{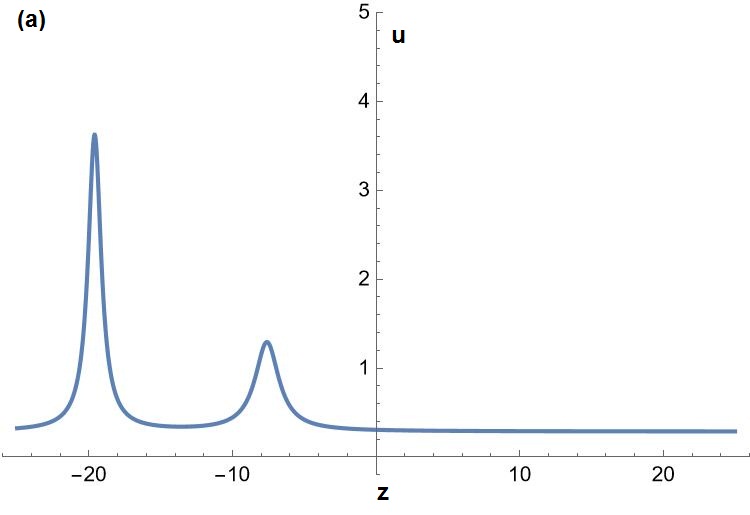}
\includegraphics[width=0.4\linewidth]{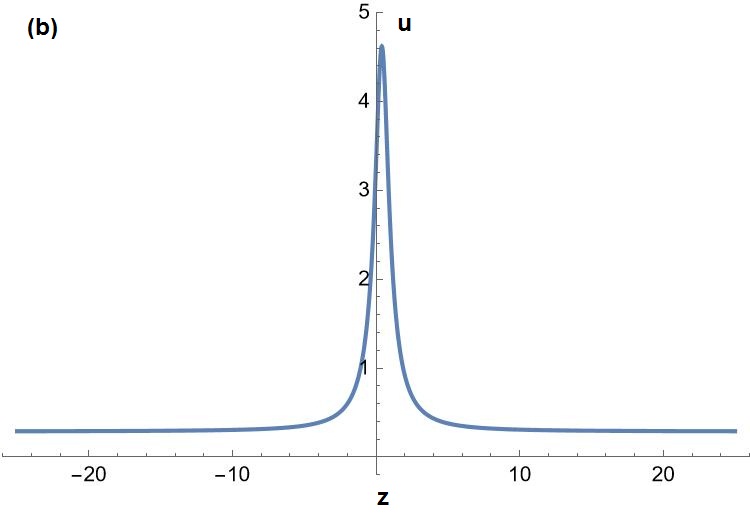}
\includegraphics[width=0.4\linewidth]{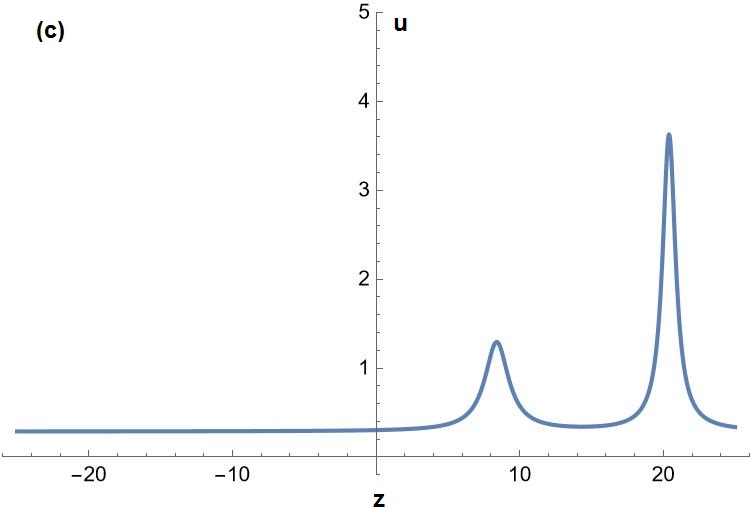}
\caption{Time evolution of two line lump solution along the same direction for time (a) t=-4, (b) t=0 and (c) t=4}
\label{2lump-same dirz}
\end{center}
\end{figure}

The time evolution of two line lumps for another set of parametric choice $x = 1; k = 0; k_1 = k_2 = 0.4; m_1=-2;m_2=-5; h = 0; q_1=1; q_2=0.3$ is shown in Fig.\ref{2lump-same dir}. For this set of parameters, the line lumps travel along the same direction along the z-axis.
The contour plot  of the above interaction is shown in Fig. Fig.\ref{2lump-same dir-cont}.
while the corresponding  two dimensional profile  in the z-direction is shown in Fig.\ref{2lump-same dirz}. Again, the line lumps undergo elastic collision.
\subsection{Hybrid Dromion}
To construct  a Hybrid dromion solution using Eq.(\ref{sol}), we choose
\begin{equation}
u_1=\mbox{sech}(z + t)^2,
\end{equation}
\begin{equation}
\xi=\mbox{sinh}y^2
\end{equation}
and substitute the above choice of arbitrary functions  in Eq.(\ref{sol}) to  generate 
\begin{equation}
    u= \mbox{sech}(t + z)^2 + \frac{1}{x + \mbox{sinh}y^2}.
\end{equation}

\begin{figure}[h!]
\begin{center}
\includegraphics[width=0.4\linewidth]{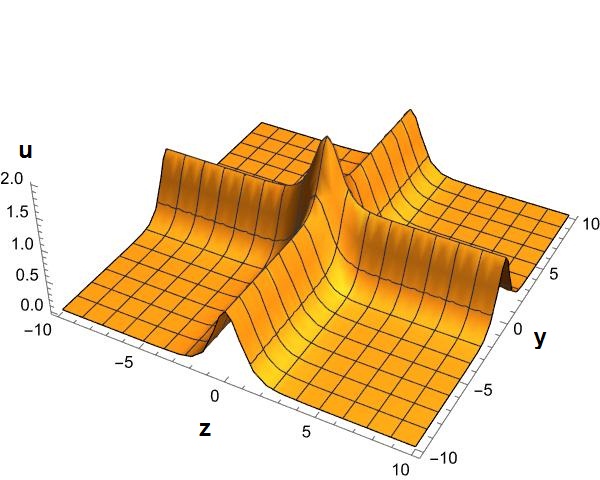}
\includegraphics[width=0.3\linewidth]{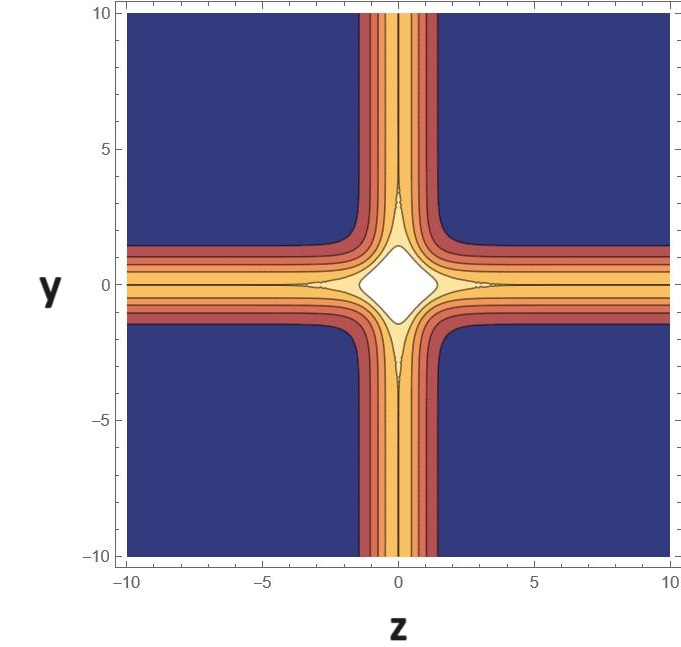}
\caption{Hybrid dromion solution and the corresponding contour plot}
\label{onedr}
\end{center}
\end{figure}

For $x = 1; t=0$, we obtain the hybrid dromion solution as shown in Fig.(\ref{onedr}). As time progresses, the dromion also travels along the z-direction. It is interesting to note that one also observes  the two nonparallel ghost solitons whose intersection generates the dromion and hence we call this solution as "Hybrid Dromion".  

\subsection{Two Hybrid Dromions}
To construct a two hybrid dromion solution using Eq.(\ref{sol}), we choose
\begin{equation}
u_1=a_1\mbox{sech}(b_1 z + c_1 t)^2+a_2 \mbox{sech}(b_2 z +c_2 t)^2,
\end{equation}
\begin{equation}
\xi=\mbox{sinh}y^2
\end{equation}
and substitute the above choice in Eq.(\ref{sol}) to generate
\begin{equation}
    u= a_1\mbox{sech}(b_1 z + c_1 t)^2+a_2 \mbox{sech}(b_2 z +c_2 t)^2 + \frac{1}{x + \mbox{sinh}y^2}.
\end{equation}

\begin{figure}[h!]
\begin{center}
\includegraphics[width=0.3\linewidth]{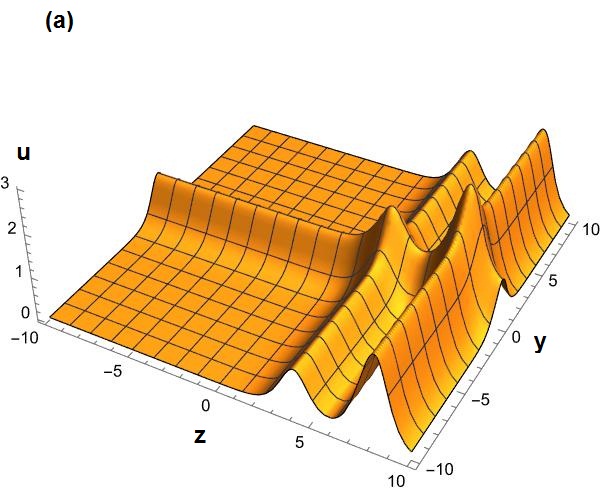}
\includegraphics[width=0.3\linewidth]{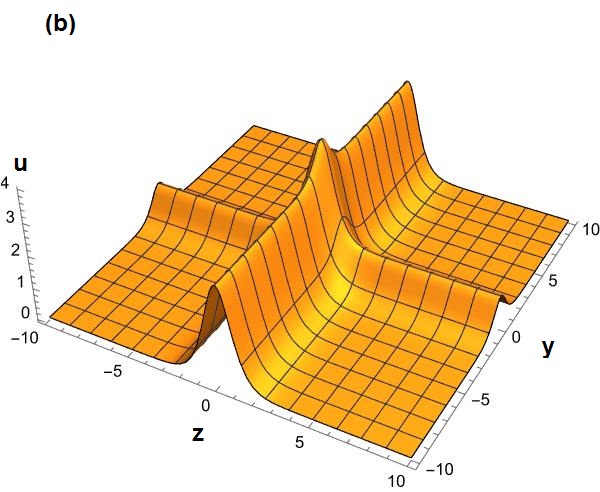}
\includegraphics[width=0.3\linewidth]{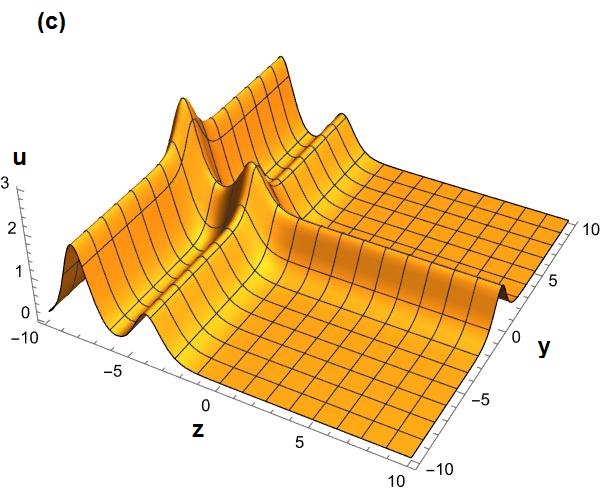}
\caption{Collisional dynamics of hybrid dromion solution for time a) t=-4, b)t=0 and c) t=4}
\label{twodr}
\end{center}
\end{figure}

\begin{figure}[h!]
\begin{center}
\includegraphics[width=0.3\linewidth]{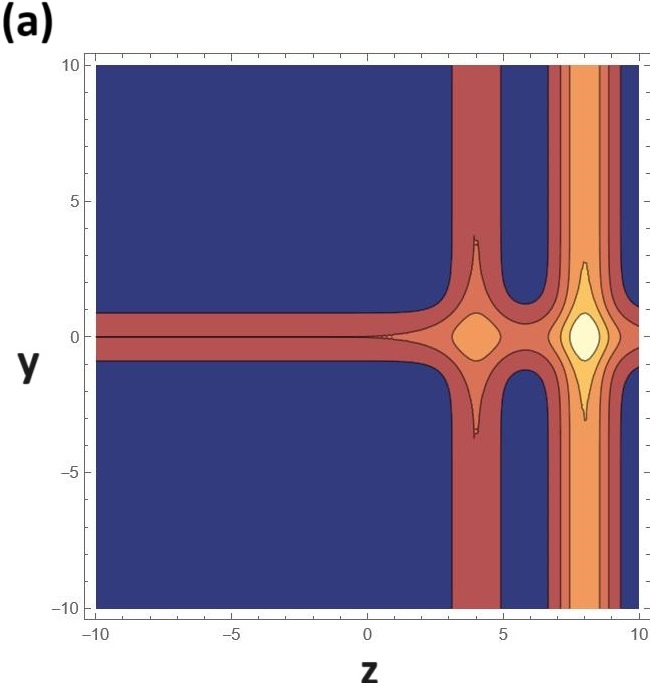}
\includegraphics[width=0.3\linewidth]{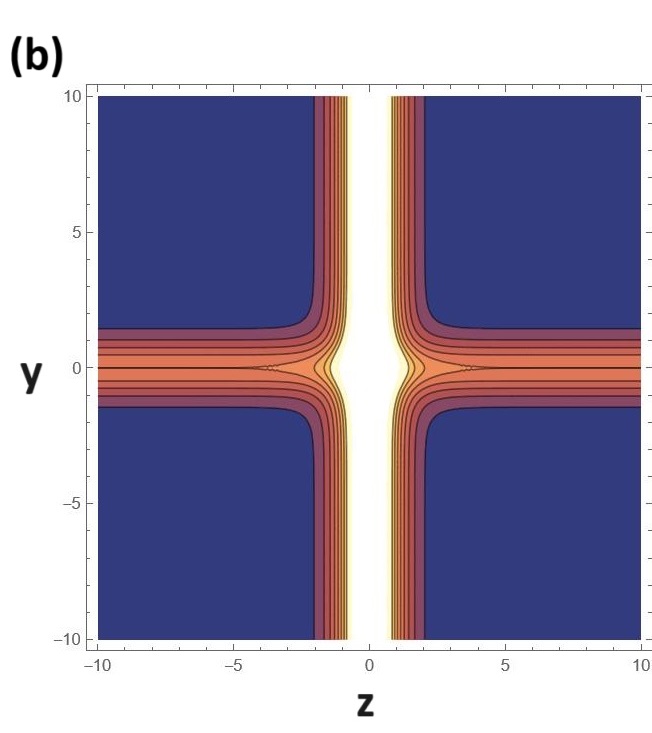}
\includegraphics[width=0.3\linewidth]{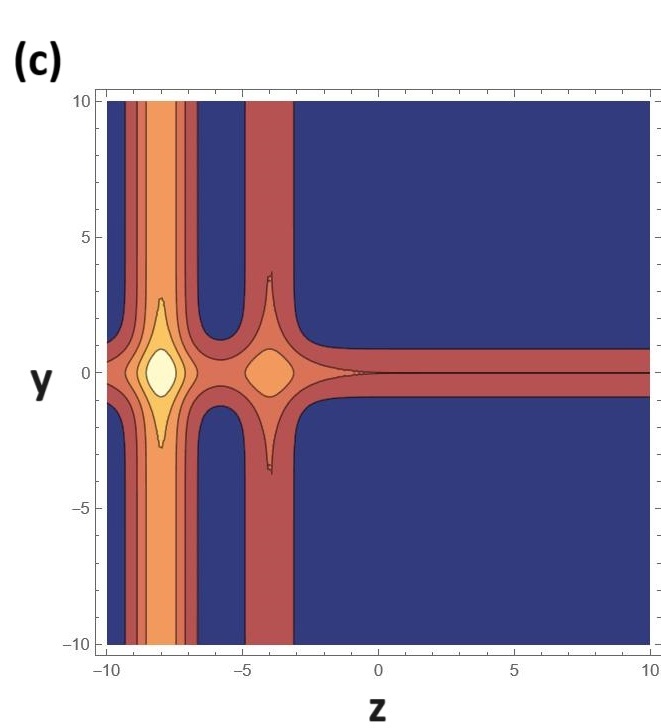}
\caption{Contour Plot of the collisional dynamics of  hybrid dromions  }
\label{twodr-cont}
\end{center}
\end{figure}

For the parametric choice $a_1=b_1=b_2=c_1=1; a_2=c_2=2; x = 1$, we obtain the two hybrid dromion solution as shown in Fig.(\ref{twodr}). One of the dromion is taller and the other one is shorter. The taller dromion travels faster than the shorter dromion.  Both  travel along the z-direction. At time t=0, they interact with each other.  After interaction, the taller dromion moves faster and overtakes the shorter one. During interaction, there is no change in amplitude of the dromions again underscoring the elastic nature of their interaction.  
The corresponding contour plots of the two hybrid dromion solution is shown in Fig. (\ref{twodr-cont}).

\section{Conclusion}
In this paper, we have investigated the (3+1) dimensional Bogoyavlensky - Konopelchenko Equation using Painlev\'e Truncation Approach and constructed  its solutions in a closed form  in terms of lower dimensional arbitrary functions of space and time. The lower dimensional arbitrary functions can be selectively chosen to  generate interesting classes of solutions like periodic waves, kinks, linear rogue waves, line lumps, dipole lumps  and hybrid dromions. We have constructed  two line lump solution and confirmed  their elastic interaction through asymptotic analysis.  Finally, we have also constructed hybrid  dromions and studied their collisional dynamics. which was again found to be elastic similar to lumps. The highlight of the results is that one can clearly observe the two nonparallel ghost solitons (in the profile of dromions) which are generally invisible in (2+1) dimensions. In addition, the lumps have been found to interact unlike in (2+1) dimensions.

From the above class of solutions, it can be emphasized that the nonlinear character of the wave has contributed to the richness in the dynamics of the solutions unlike linear waves which disperse and fizzle out during time evolution.
We do believe that the above analysis employing Truncated Painlev\'e approach can be effectively used to study  physically important higher dimensional systems to construct more general solutions exhibiting  interesting collisional dynamics which have not been witnessed so far.

\bibliographystyle{elsarticle-num-names}

\bibliography{cas-refs}

\end{document}